\documentclass[lettersize,journal]{IEEEtran}

\usepackage{subfigure}
\usepackage{amsmath,amsfonts}
\usepackage{algorithmic}
\usepackage{algorithm}
\usepackage{array}
\usepackage{textcomp}
\usepackage{stfloats}
\usepackage{url}
\usepackage{verbatim}
\usepackage{graphicx}
\usepackage[compress]{cite} 
\usepackage{float}
\usepackage{booktabs}
\usepackage{multirow}
\usepackage{makecell}
\usepackage{amssymb,amsmath,amsfonts}
\usepackage{mathtools}
\usepackage{color}
\usepackage[table,xcdraw]{xcolor}
\usepackage{booktabs}
\usepackage{bbding}

\usepackage{tikz,xcolor}
\usepackage[colorlinks,linkcolor=blue]{hyperref}
\hypersetup{hidelinks,
	colorlinks=true,
	allcolors=black,
	pdfstartview=Fit,
	breaklinks=true}
\definecolor{lime}{HTML}{A6CE39}
\DeclareRobustCommand{\orcidicon}{
\begin{tikzpicture}
\draw[lime, fill=lime] (0,0)
circle[radius=0.16]
node[white]{{\fontfamily{qag}\selectfont \tiny \.{I}D}};
\end{tikzpicture}
\hspace{-2mm}
}
\foreach \x in {Ronghao,Qiaolin,Sijie,Zefeng,Huhaifeng,Tanyappeng,A, ..., Z}{%
\expandafter\xdef\csname orcid\x\endcsname{\noexpand\href{https://orcid.org/\csname orcidauthor\x\endcsname}{\noexpand\orcidicon}}
}

\title{Divide and Conquer: Mixture-of-Bottleneck Experts in Informative Ordinal Space for Video-based Multimodal Sentiment Analysis}

\author{Ronghao~Lin\hspace{-1.0mm}\orcidRonghao{}\hspace{-1.0mm},
Qiaolin~He,
Zefeng~Lu\hspace{-1.0mm}\orcidZefeng{}\hspace{-1.0mm},
Yichu~Liu,
Li~Huang,
Sijie~Mai\hspace{-1.0mm}\orcidSijie{}\hspace{-1.0mm},
Haifeng~Hu\hspace{-1.0mm}\orcidHuhaifeng{} \hspace{-2.0mm}\IEEEmembership{,~Member,~IEEE
}, 
and Yap-peng~Tan\hspace{-1.0mm}\orcidTanyappeng{} \hspace{-2.0mm}\IEEEmembership{,~Fellow,~IEEE
}

\thanks{Ronghao Lin is with College of Computer Science and Software Engineering, Shenzhen University, Shenzhen 518060, China.  (E-mail: linrh@szu.edu.cn).}
\thanks{Qiaolin He, Li Huang, and Haifeng Hu are with the School of Electronics and Information Technology, Sun Yat-sen University, Guangzhou 510006, China (E-mail: \{heqlin5,huangli63\}@mail2.sysu.edu.cn, huhaif@mail.sysu.edu.cn).}
\thanks{Zefeng Lu is with School of Cyberspace Security, Guangzhou University, Guangzhou 510006, China (E-mail: luzefeng@gzhu.edu.cn).}
\thanks{Sijie Mai is with School of Computer Science, South China Normal University, Guangzhou 510631, China (E-mail: sijiemai@m.scnu.edu.cn).}
\thanks{Yap-peng Tan is with VinUniversity, Hanoi, Vietnam and Nanyang Technological University, Singapore, 639798 (E-mail:yp.t@vinuni.edu.vn).}
\thanks{Ronghao Lin is also with School of Electrical and Electronic Engineering, Nanyang Technological University, Singapore, 639798.}
\thanks{Yichu Liu, Li Huang are with Desay SV Automotive Co., Ltd, Huizhou, Guangdong, China (E-mail: {yichu.liu,Li.Huang}@desaysv.com)}

}

\IEEEpubid{0000--0000/00\$00.00~\copyright~2021 IEEE}

\begin{document}

\maketitle

\begin{abstract}
Video-based Multimodal sentiment analysis (MSA) must handle information from text, audio, and image sequence in human speaking videos, yet current methods often fail to integrate modalities with task awareness. Most models treat video sentiment prediction as a single task, overlooking its ordinal nature, and their fusion strategies struggle to capture diverse unique and synergic cues across modalities. To address these limitations, we adopt a divide-and-conquer perspective by reformulating MSA as an ordinal regression problem and decoupling it into polarity recognition and intensity prediction. Driven by information theory, we introduce a \textbf{Mixture-of-Bottleneck (MoB)} framework that assigns different latents to polarity- and intensity-specific experts for different modalities. With the learning of information bottleneck, each expert learns compact and task-relevant representations while filtering out redundancy and noise. A multimodal bottleneck routing fusion module then fuses these expert latents with hard mining strategy, guiding the prediction in the ordinal sentiment space. Extensive experiments on 4 MSA datasets and 4 language models show that MoB effectively leverages informative latents from diverse modalities and captures general sentiment structure. Beyond stronger performance, MoB comprehensively captures fine-grained intra- and inter-modal dynamics, enabling more trustworthy localization of nuanced video sentiment signals. 
\end{abstract}


\section{Introduction}
\IEEEPARstart{M}{ultimodal} machine learning \cite{liang2024foundations} aims to integrate and reason over heterogeneous data sources such as text, audio, and image. As a sub-field of multimodal learning, video-based Multimodal Sentiment Analysis (MSA) aims at jointly conveying rich affective cues from the human speaking videos,
 where spoken words provide explicit semantics, tone reflects emotional strength, and facial or bodily movements reveal implicit attitude \cite{pandey2024progress}. However, effectively combining these heterogeneous sources remains challenging due to modality heterogeneity, task diversity, and the ordinal nature of sentiment intensity \cite{poria2020beneath,stoehr2023sentiment}.

\begin{figure}[htbp]
\centering 
\includegraphics[scale=0.55]{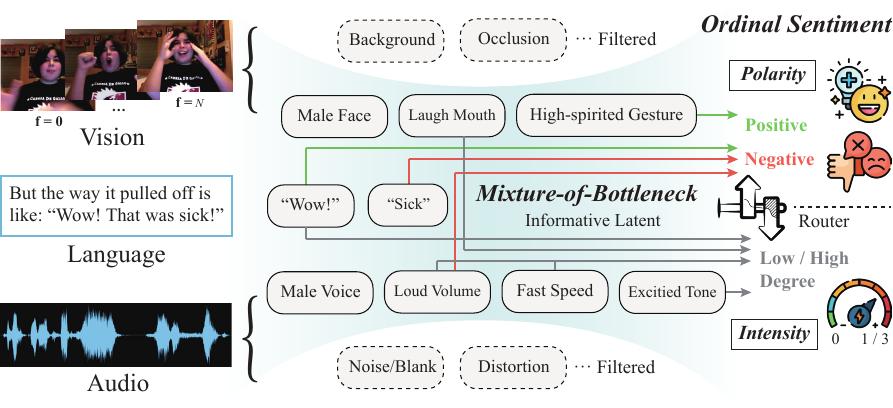}
\caption{Illustration of task-related information extraction for each modality in the ordinal sentiment space for video-based multimodal sentiment analysis. }
\label{fig_intro}
\vspace{-0.5cm}
\end{figure}

Current MSA approaches mainly focus on building strong unimodal encoders and designing exquisite multimodal fusion strategies \cite{gkoumas2021makes}. Despite recent advances, most fusion mechanisms remain task-agnostic and combine features from different modalities in a static manner \cite{han2022dynamics,xue2023dynamic}. Besides, they typically treat sentiment prediction as a single task, ignoring its intrinsic hierarchical and ordinal structure \cite{xiao2024neuroinspired,wang2025dlf}. These limitations prevent previous methods from considering the distinctive roles that different modalities play in recognizing diverse aspects of sentiment. Consequently, they are easily affected by unimodal noise or cross-modal redundancy, leading to reduced interpretability and sub-optimal performance \cite{lin2024semanticmac}. 
As illustrated in the example of Figure \ref{fig_intro}, visual gestures may better convey polarity (positive vs. negative), while acoustic tone more effectively captures intensity (emotional strength), which emphasizes the need for dynamic and task-aware multimodal learning in video-based MSA.

\IEEEpubidadjcol

To address these challenges, we rethink MSA from an information-theoretic and divide-and-conquer perspective. The Information Bottleneck (IB) principle provides a principled way to extract task-relevant yet compact latent representations by filtering out redundant or noisy signals \cite{tishby2015deepib}. However, conventional IB-based models treat all features equally and fail to capture the varying difficulty and complementary nature of modal-specific and -shared feature \cite{liang2023factorized,mai2023mib,liang2023pid,xiao2024neuroinspired}. Meanwhile, Mixture-of-Experts (MoE) networks \cite{noam2017outrageously} specialize experts to divide and process modality-specific information in multimodal learning \cite{yu2024mmoe,han2024fusemoe}, yet they are typically limited to feature-level routing and lack task-level disentanglement.

Motivated by these observations, we propose the \textbf{Mixture-of-Bottleneck (MoB)} framework, as illustrated in Figure~\ref{fig_intro}, which unifies the Information Bottleneck principle and Mixture-of-Experts routing as a dynamic multimodal learning scheme. We first reformulate MSA task in an ordinal sentiment space and explicitly decouple it into polarity recognition and intensity estimation. This division aligns with the intrinsic structure of sentiment and enables each subtask to exploit the relevant cues from each modality. MoB then assigns polarity and intensity bottleneck experts to process the unimodal informative latents extracted by IB, capturing minimal yet sufficient information for each subtask. Moreover, we present a multimodal bottleneck router to adaptively fuse the unimodal latents and produce the final sentiment prediction with an effective hard mining strategy in the ordinal space. Following a divide-and-conquer paradigm, MoB partitions the feature space into multiple informative bottleneck experts and the label space into ordered polarity and intensity levels, achieving both interpretability and adaptability in multimodal fusion for video-based MSA. Our contribution can be summarized as:

\begin{itemize}
\item \textbf{Ordinal Sentiment Modeling:} We rethink video-based MSA as an ordinal regression task, decoupling polarity recognition and intensity estimation to better reflect the natural hierarchical structure of sentiment.
\item \textbf{Mixture-of-Bottleneck Experts:} The MoB framework integrates the Information Bottleneck principle with expert specialization, assigning polarity- and intensity-specific bottleneck experts to adaptively extract compact and task-relevant latents from each modality.
\item \textbf{Dynamic Bottleneck Routing Fusion:} A multimodal bottleneck router dynamically fuses task-aware unimodal latents across modalities, enabling fine-grained inter-modal interaction and capturing diverse cross-modal synergy with various combination of information.
\item \textbf{Extensive and Comprehensive Experiment:} Comprehensive experiments on 4 datasets and 4 pre-trained language models demonstrate that MoB consistently outperforms state-of-the-art methods, while offering enhanced interpretability for video-based MSA.
\end{itemize}

\section{Related Work}
\subsection{Multimodal Sentiment Analysis}
Video-based Multimodal Sentiment Analysis (MSA) has emerged as a critical research area in cognition computation, aiming to identify speakers' sentiments by integrating information from multiple modalities in the talking videos, such as text, audio, and image modality signals \cite{poria2020beneath,pandey2024progress}. 
Previous approaches primarily focused on capturing unimodal semantic representations and designing effective multimodal fusion strategies \cite{gkoumas2021makes}, which can be broadly categorized into early \cite{zadeh2018multi,yang2023confede}, late \cite{yu2021learning,han2021improving, hu2022unimse,wu2024mmml}, and hybrid fusion methods \cite{ lin2022mmcl, lin2023mtmd,zhang2023learning,sun2024hkdmer,zhang2025modal}. Since the fusion process is typically task-agnostic, these methods fail to provide interpretability for sentiment prediction, resulting in static multimodal fusion \cite{han2022dynamics,xue2023dynamic}.
Besides, considering the semantic dependency nature of the MSA task \cite{lin2024semanticmac}, performance improvement increasingly rely on the pre-trained language models \cite{hu2022unimse, lin2023dynamically, zhang2023learning}. 
However, due to the heterogeneity issues \cite{ zhang2025modal}, integrating textual features \cite{devlin2019bert,liu2019roberta,he2023debertav3}with acoustic and visual cues from speaker videos remains a major challenge. Thus, developing unified  multimodal frameworks that can effectively combine video-based signals with diverse pre-trained language models are imperative in the field of MSA \cite{lin2024semanticmac}.

\subsection{Information-Theoretic Deep Learning}
Information theory has been widely utilized in the development of deep learning models, offering a interpretable framework to conduct representation learning for understanding various modalities \cite{liang2023factorized,liang2023pid}. 
Information Bottleneck (IB)  \cite{tishby1999information,tishby2015deepib} has been extended and adapted in various multimodal contexts, beneficial in filtering task-irrelevant noise and capturing intra-modal dynamics in MSA domain \cite{mai2023mib,xie2024trustworthy,guo2024embracing,xiao2024neuroinspired}. Regardless of the success in learning task-relevant features, IB remains limited in disentangling modality-specific and -shared representations, which is crucial for efficient multimodal representation learning \cite{liang2023pid}. 
Besides, existing methods typically apply VIB to individual tasks separately, without considering efficient fusion or coordination across various bottlenecks, especially in scenarios where broader multimodal learning goal accompanied by diverse unimodal learning biases \cite{zhang2024understanding}. Such gap highlights the motivation of our method in discovering more dynamic approaches for IB-based multimodal representation learning.

\subsection{Mixture-of-Expert Network}
Mixture-of-Experts (MoE) was originally introduced as a conditional computation strategy in which a sparse gating network dynamically routes each input to a subset of specialized experts \cite{noam2017outrageously,zhou2022choice}.
MoE has been extended to multimodal learning where routing networks choose experts specialized for different modalities or modality combinations recently \cite{yu2024mmoe,lin2026moellava}. 
Due to the productive effect in capturing relationships across diverse tasks, We consider MoE as a compelling mechanism in enhancing the performance and flexibility of information bottleneck learning. By embedding unimodal bottleneck with multiple experts, we can dynamically compress and store information based on the difficulty and relevance of specific subtasks.

\subsection{Ordinal Nature of Sentiment}
Modeling sentiment has been a long-standing and challenging research due to the inherent subjective and ambiguous nature of sentiment cues \cite{yannakakis2021ordinal,stoehr2023sentiment}. 
Recent advances in ordinal sentiment learning \cite{xie2024trustworthy,tellamekala2024coldfusion} have demonstrated the potency of treating affective labels as inherently relative rather than absolute. 
Inspired by the separation thought of sentiment aspects \cite{tian2018polarity}, we decompose the sentiment space into distinctive polarity and intensity dimensions, thereby decoupling the multimodal learning process of sentiment modeling. By doing so, the model can adaptively assign the related polarity and intensity cues of each modality and exploit ordinal information at both label and feature levels.

\section{Methodology}


Considering MSA task in a divide-and-conquer paradigm, we firstly construct the ordinal sentiment space in Sec. \ref{section_ordinal_sentiment_space}, which divide the original sentiment space into two decoupled subspace to explicitly enable task-aware feature extraction. Then, we present PIBE in Sec. \ref{section_pi_bottleneck_expert}, aiming to combine the strength of MoE and IB in dynamically integrating task-related information and filtering task-unrelated noise according to the supervision from each subspace. Next, we present MBRF in Sec. \ref{section_multimodal_bottleneck_routing_fusion}, focusing on adaptively capturing the cross-modal dynamics and exploring the optimal way in conducting multimodal information fusion. Lastly, we design a hard mining strategy based on the ordinal sentiment space to enhance convergence on difficult samples and present the total multi-task optimization objective in Sec. \ref{section_multimodal_bottleneck_routing_fusion}.

\begin{figure*}[htbp]
\centering 
\includegraphics[scale=0.65]{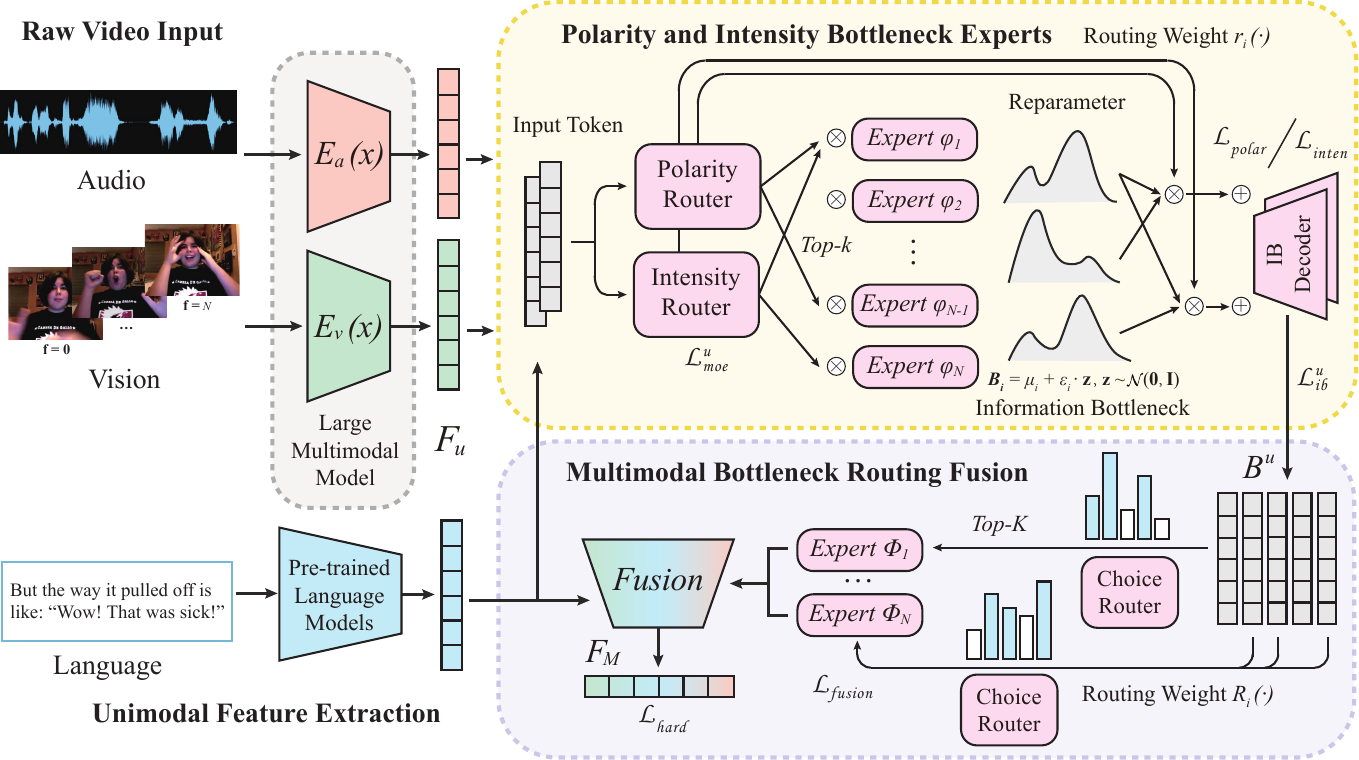}
\caption{The framework of Mixture-of-Bottleneck (MoB) to dynamically process modalities in the informative ordinal space. }
\label{fig_architecture}
\end{figure*}

\subsection{Problem Definition}
Given human-centric videos with three modalities, including language modality $X_l$ (utterance), acoustic modality $X_a$ (speaking audios), and visual modality $X_v$ (face frames), the task of multimodal sentiment analysis aims at extracting features $F_u\in\mathbb{R}^{N_u\times d_u}$ for unimodal signals $u\in\{l,a,v\}$ to conduct multimodal fusion and leveraging the multimodal representation $F_M\in\mathbb{R}^{d_M}$ analyze the sentiment state $\hat{y}\in\mathbb{R}^{\mathbb{1}}$ of the speaker in the video. Note that $N_u$ denotes the sequence length for temporal unimodal data in the video, while $d_u$ and $d_M$ denotes the feature dimension. With the ground truth sentiment labels $y$, the optimization objective for the final sentiment analysis task can be formulated as:
\begin{equation}
    \mathcal{L}_{task}=MSE(y,\hat{y})=\parallel y-\hat{y}\parallel^2
\end{equation}

\subsection{Ordinal Sentiment Space}
\label{section_ordinal_sentiment_space}
Since sentiment analysis task demands of abundant semantics from diverse modalities, directly predicting the final sentiment states may raise issues of coarse affective understanding and subjective unimodal bias \cite{poria2020beneath}. Inspired by previous ordinal learning methods \cite{xie2024trustworthy,tellamekala2024coldfusion}, we organize the ground truth sentiment space in ordinal regression manner by dividing the sentiment state into two fine-grained aspects, polarity and intensity, referring to positive/neutral/negative tendency and the strength degree of sentiment \cite{tian2018polarity}, defined as:
\begin{equation}
    y_{polar}=\left\{
\begin{aligned}
    & +1,\ \ y>0\\
    & 0,\ \ y=0\\
    & -1,\ \ y<0
\end{aligned}
\right.
\label{equ_y_polar}
\end{equation}
\begin{equation}
    y_{inten}=\mid y\mid < H, \ \ \forall y
\label{equ_y_inten}
\end{equation}
where $y_{polar}$ and $y_{inten}$ denote the polarity and intensity ground truth labels, $H$ denote the widest annotation range for diverse datasets. 

\subsection{Polarity and Intensity Bottleneck Experts}
\label{section_pi_bottleneck_expert}
Considering modality input $X_u$, where $u\in\{l,a,v\}$ denote language, acoustic, and visual modality. After unimodal extraction by pre-trained large multimodal model, the feature of each modality can be denotes as $F_u\in\mathbb{R}^{n_u\times d_u}$ where $n_u$ denotes the sequence length for temporal unimodal data in the video, while $d_u$ denotes the feature dimension. To sufficiently capture task-related information, we leverage Information Bottleneck (IB) to extract unimodal features by variation approximation \cite{tishby1999information,tishby2015deepib, alemi2017vib} with the supervision of diverse targets. 

Given the input $S$ and target $T$ to construct the Markov chain as $S\rightarrow B\rightarrow T$, the optimization of mutual information learning can be formulated as:
\begin{equation}
    \min \mathcal{L}_{IB}(S,T) = \min_{p(B|S)} I(S;B) - \beta\ I(B;T)
\label{equ_ib}
\end{equation}
where $B$ denotes the bottleneck containing necessary information of input $S$ to conduct the task prediction for $T$ and $\beta$ serves as trade-off for two mutual information terms.

To inherit the optimization ability for backward gradient propagation of neural networks, previous methods \cite{kingma2014vae,alemi2017vib} adopt variational approximation $q(T|B)$ for the conditional probability in Equ. \ref{equ_ib} and turn the optimization of IB into:
\begin{equation}
 \min \mathbb{E}_S \Big[KL( p(B|S)\parallel q(B))\Big] - \beta\ \mathbb{E}_{B,S} \Big[\log q(T|B)\Big]
\label{equ_ib2}
\end{equation} 

Then a reparameterization trick utilizes multi-layer perception ($MLP$) as the IB encoder to output the mean and variance for the bottleneck of the informative latents, which can be formulated as:
\begin{equation}
    B \sim\mathcal{N}(\mu, \varepsilon^2), \text{where } \{\mu, \varepsilon\} =MLP(S)
\end{equation}
where $B=\mu+\varepsilon\cdot \mathbf{z}$ and $\mathbf{z}\sim\mathcal{N}(0, \mathbf{I})$ denotes to sample from a standard Gaussian distribution. 
Then for Equ. \ref{equ_ib2}, the first term can be derived into $\mathcal{L}_{ib}=KL(\mathcal{N}(\mu, \varepsilon^2)\parallel\mathcal{N}(0, \mathbf{I}))$ while the second term denotes the task prediction loss.



Aiming at adaptively extracting both unique and shared information from unimodal features and assigning various task-related bottlenecks to the downstream subtasks, we then utilize Mixture-of-Expert (MoE) networks \cite{noam2017outrageously,riquelme2021scaling} to replace the original $MLP$ as the IB encoders. 
For $F_u$, the bottleneck expert consists of a group of $N$ experts network $\{\varphi_1,\varphi_2,...,\varphi_{N-1},\varphi_{N}\}$ along with a router $r(\cdot)$ which dynamically determine the flowing of each information bottleneck. Then, the output $\widetilde{y}$ of the Mixture-of-Bottleneck Experts network is derived from the summation of Top-$k$ related expert, denoted as:
\begin{equation}
\begin{aligned}    \widetilde{y}&=\sum_{i=1}^{N}r(B)_i\cdot\varphi_i(B) \\
    &= \sum_{i=1}^{N}r(B)_i\cdot \mathcal{N}(\mu;\mu_i, \varepsilon^2_i)
\end{aligned}
\label{equ_mob}
\end{equation}
where $r(B)_i=0$ when $i$ is not in the top $k$ elements and the output can be regarded as a special type of Gaussian Mixture Models \cite{chen2025gmflow} with adaptive weights rely on the input.

The input latent $B$ from Equ. \ref{equ_mob} consists of task-related features from each modality while does not contain task-aware information. Inspired by \cite{liang2022m3vit,chen2023adamvmoe}, we employ two settings to feed the subtask guidance into the router:

\noindent\textbf{Multi-gate Setting.} The router utilizes $MLP$ as non-linear projection with $GELU$ activation layer \cite{hendrycks2016gelu} for each subtask to specify the corresponding experts, denoted as:
\begin{equation}
    r_{task}(B)=Gate(MLP(B;\theta_{task}))\in\mathbb{R}^{b\times N}
\end{equation}

\noindent\textbf{Multi-embed Setting.} The information of each subtask is initialized with a specific embedding $Emb_{task}$ and then added into the original input for the router, denoted as:
\begin{equation}
    r_{task}(B)=Gate(B+Emb_{task})\in\mathbb{R}^{b\times N}
\end{equation}
where $b$ denotes the batch size and $N$ denotes the number of bottleneck experts.

By these settings, the router can adaptively decide the experts with specific or sharing weights for different subtasks based on the $Gate$ function. Following \cite{noam2017outrageously}, both tunable Gaussian noise and additional soft constraint are included to enhance the load balance when routing bottleneck experts. The balancing constraint can be denoted as:
\begin{equation}
    \mathcal{L}_{moe}=CV^2\Big[\sum\nolimits_{B} r(B)\Big] + CV^2\Big[\sum\nolimits_{B} P(B,i)\Big]
\label{equ_moe}
\end{equation}
\begin{equation}
    \text{where } CV^2(x)=\left\{
    \begin{aligned}
    & 0 & \text{if } |x| = 1, \\
    & \dfrac{\operatorname{Var}(x)}{\left(\operatorname{E}[x]\right)^2 + 1e^{-10}} & \text{otherwise.}
    \end{aligned}
\right.\ \ 
\end{equation}
where $P(B,i)$ is defined as the probability that $r(B)_i$ is non-zero when we resample the noise on the $i$-th expert while keeping all previously sampled noise on the other elements fixed. With such constraint, the router tend to encourage similar numbers of bottlenecks and strengthen the  convergence efficiency of the model.

The IB decoder is lastly adopted to map the output of Mixture-of-Bottleneck Experts into the sentiment label space. Considering ordinal learning by polarity and intensity prediction of sentiment, the subtasks prediction loss for each router is defined as follows:

\noindent\textbf{Polarity-guided Router.} With the supervision label $y_{polar}$ from Equ. \ref{equ_y_polar}, the objective for polarity-guided router is computed in a classification manner, denoted as:
\begin{equation}
    \mathcal{L}_{polar}=-\mathbb{E}_x\
    \Big[y_{polar}\log\widetilde{y}_{polar}\Big]
\label{equ_polar}
\end{equation}

\noindent\textbf{Intensity-guided Router.} With the supervision signal $y_{inten}$ from Equ. \ref{equ_y_inten}, the objective of intensity-guided router is measured by regression learning, denoted as:
\begin{equation}
    \mathcal{L}_{inten}=\mathbb{E}_x\ \Big[\big|y_{inten} - \widetilde{y}_{inten}\big|\Big]
\label{equ_inten}
\end{equation}

For unimodal features $u\in\{l,a,v\}$, deriving from Equ. \ref{equ_ib2}, \ref{equ_moe},\ref{equ_polar} and \ref{equ_inten}, the optimization objective of polarity and intensity bottleneck experts is denoted as:
\begin{equation}
    \mathcal{L}_{btnk}= \sum\nolimits_{u}(w_{ib}\mathcal{L}_{ib}^u +w_{moe} \mathcal{L}_{moe}^u)+w_{ord}(\mathcal{L}_{polar}+\mathcal{L}_{inten})
\end{equation}
where $w_{ib}$, $w_{moe}$ and $w_{ord}$ are hyper-parameters adjusting the contribution of IB compression, MoE load balancing and subtask prediction.

\subsection{Multimodal Bottleneck Routing Fusion}
\label{section_multimodal_bottleneck_routing_fusion}
To capture unique and shared information inside and across modalities, we propose Multimodal Bottleneck Routing Fusion (MBRF) to explore various cross-modal synergy with diverse combinations of unimodal bottlenecks.

Given the unimodal informative latents $\{B^u_j|_{j=1}^b\}\in\mathbb{R}^{n_b\times d}$ extracted by previous bottleneck experts and a series of fusion experts $\{\phi_1,...,\phi_{N}\}$ , we can derive a latent-to-expert affinity score matrix $S_j$ for each instance:
\begin{equation}
    S(B)_j=Gate(B^u_j)\in\mathbb{R}^{6\times N}
\end{equation}
where $N$ denotes the number of experts and $d$ is the common hidden dimension and $n_b=b\times 6$ denotes the total number of latents for each batch with batch size $b$, since each modality contains two kinds of informative latents output by polarity and intensity router, respectively.

Then we can compute the routing weight matrix $\mathcal{R}$(B) of expert for the selected latents and the index matrix $\mathcal{I}$ with selected latents, denoted as:
\begin{equation}
    \mathcal{R}(B), \mathcal{I}=TopK(S^\top,K)
\end{equation}
where $\mathcal{I}=(I[i,j])_{N\times K}$ specifies $j$-th selected latent of the $i$-th expert. The aggregated output of all fusion experts $\phi_i$ can be denoted as: 
\begin{equation}
    \Delta_j=\sum_{i,j\in\mathcal{I}_i}\mathcal{R}(B)_i\cdot\phi_i(B)
\end{equation}

Considering enhancing such cross-modal dynamics for each unimodal latents $F_u$ \cite{xiao2024neuroinspired}, we adopt residual connection \cite{he2016resnet} to obtain the final multimodal representation as:
\begin{equation}
    F_M=Concat(F_l,F_a,F_v)+\Delta \in\mathbb{R}^{b\times6d}
\end{equation}

Similar with \cite{zhou2022choice}, we regularize the multimodal routing fusion process by limiting the maximization number of experts for each latent. We firstly apply Softmax over the transposed latent-to-expert score matrix $S$ to produce the probabilistic routing distribution where each expert softly selects latents according to their relative scores, denoted as:
\begin{equation}
A_{ij} = \frac{\exp(S^\top_{ij})}{\sum_{m=1}^{n} \exp(S^\top_{im})}
\label{eq_soft_assignment}
\end{equation}


Besides, we add an entropy regularization term for matrix $A$ to discourages overly peaked assignments by penalizing low-entropy distributions, denoted as:
\begin{equation}
\mathcal{L}_{et} = - \sum_{i=1}^{N} \mathbb{E}_j\Big[ A_{ij} \log A_{ij}\Big]
\label{eq:entropy}
\end{equation}

Combining both the constraint penalties and the entropy term, the fusion routing objective can be denoted as a entropy-regularized linear programming problem:
\begin{equation}
\mathcal{L}_{fusion} =
- \sum_{i=1}^{N} \mathbb{E}_j\Big[ S^\top_{ij} A_{ij}\Big]
+ \lambda\cdot\mathcal{L}_{et}\ ,
\ \forall j: \sum_{i=1}^{N} A_{ij} \leq h
\label{eq:fusion_loss}
\end{equation}
where each latent is routed to at most $h$ experts.


\subsection{Multi-task Objectives}
To further explore the ordinal sentiment space, we adopt a hard-sample mining strategy that dynamically adjusts the task loss based on the polarity and intensity prediction errors $\delta$. We define three types of hard samples $C_h$ selected with the prediction errors $\delta$:
\begin{equation}
\begin{aligned}
C_1 &:~ \delta_p = 1,~ \delta_i \le \tau \\
C_2 &:~ \delta_p = 0,~ \delta_i \ge 2\tau \\
C_3 &:~ \delta_p = 1,~ \delta_i \ge 2\tau
\end{aligned}
\text{\ ,\  with}
\left\{
\begin{aligned}
\delta_p &= \mathbb{I}\!\left[\tiny{\arg\max(\hat{y}_{\text{polar}}) \neq y_{\text{polar}}}\right]\\
\delta_i &= \left|\hat{y}_{\text{inten}} - y_{\text{inten}}\right|
\end{aligned}
\right.
\end{equation}
where $\tau$ denotes the difficulty threshold of the sentiment score range $H$. Thus, the hard mining loss is defined as:
\begin{equation}
    \mathcal{L}_{hard}=\sum_{h=1}^3w_h\big|MLP(F_M)-y_{gt}\big|
\end{equation}
where $w_h$ denotes the weight assigned for hard sample $C_h$, which encourages the model to learn more effectively from the difficult cases.


Overall, the joint training process for above modules including bottleneck extraction, experts balancing, and hard mining, the multi-task objective can be summarized as:
\begin{equation}
\mathcal{L}_{total} = \mathcal{L}_{task}+\mathcal{L}_{btnk}+ w_{fus}\cdot\mathcal{L}_{fusion} + w_{hard}\cdot\mathcal{L}_{hard}
\label{eq:total}
\end{equation}

\section{Experiment}

\subsection{Datasets} We evaluate our method on multilingual MSA task, including 2 English and 2 Chinese corpora.
\textit{CMU-MOSI (MOSI)} \cite{zadeh2016mosi} contains 2,199 monologue utterances from 93 opinion-based YouTube videos featuring 89 movie reviewers. Each segment is annotated for sentiment on a continuous scale ranging from -3 (strongly negative) to +3 (strongly positive).
\textit{CMU-MOSEI (MOSEI)} \cite{zadeh2018mosei} extends MOSI by providing roughly 66 hours of video clips across 250 topics from 1,000 speakers with the same sentiment range. 
\textit{CH-SIMS (SIMS)} \cite{yu2020chsims} comprises 2,281 segments from 60 Chinese videos in different films, TV dramas, and variety shows, covering natural expressions with varied poses, occlusions, and illumination from 474 speakers, using the sentiment scale from -1 (negative) to +1 (postive).
\textit{CH-SIMS v2 (SIMSv2)}\cite{liu2022chsims2} approximately doubles the scale by adding more supervised and unsupervised samples under the same annotation scheme where only supervised samples are used in our papers for fair comparison.




\begin{table*}[t!]
\caption{Performance Comparison between the proposed MoB and baselines on CMU-MOSI and CMU-MOSEI datasets. The baseline models are reproduced except for $^\dag$ meaning that the corresponding results are copied from the original paper. '-/-' for Acc2 and F1 denote the two evaluation settings of non-negative/negative and positive/negative respectively. }
\label{table_mosi_mosei}
\centering
\scalebox{0.9}{
\begin{tabular}{c|ccccc|ccccc}
    \toprule[1.0pt]
    \multirow{2}{*}{Models} & \multicolumn{5}{c|}{CMU-MOSI} & \multicolumn{5}{c}{CMU-MOSEI}\\
    & Acc7$\uparrow$  & Acc2$\uparrow$ & F1$\uparrow$ & MAE$\downarrow$ & Corr$\uparrow$ & Acc7$\uparrow$  & Acc2$\uparrow$ & F1$\uparrow$ & MAE$\downarrow$ & Corr$\uparrow$\\
    \midrule[1.0pt]
    Self-MM \cite{yu2021learning} &  45.8 &  82.7 / 84.9  &  82.6 / 84.8  & 0.731  & 0.785  & 
    53.0 &  82.6 / 85.2  &  82.8 / 85.2  &  0.540 & 0.763 \\
    MMIM \cite{han2021improving} & 45.0 &  83.0 / 85.1  &  82.9 / 85.0  & 0.738  & 0.781  & 
    53.1 &  81.9 / 85.1  &  82.3 / 85.0  & 0.547  &  0.752 \\
    MMCL \cite{lin2022mmcl} & 46.5 &  84.0 / 86.3  &  83.8 / 86.2  & 0.705 & 0.797  & 53.6  &  84.8 / 85.9  &  84.8 / 85.7  & 0.537  &  0.765 \\ 
    ConFEDE \cite{yang2023confede} & 42.3 & 84.2 / 85.5  & 84.1 / 85.5  & 0.742 & 0.784 & 54.9 & 81.7 / 85.8 & 82.2 / 85.8 & 0.522 & 0.780 \\
    MIB$^\dag$ \cite{mai2023mib} & 48.2 & - / 85.2 & - / 85.2 & 0.728 & 0.793 & 53.0 & - / 86.2 & - / 86.2 & 0.584 & 0.789 \\
    MTMD \cite{lin2023mtmd} & 47.5  &  84.0 / 86.0  &  83.9 / 86.0  &  0.705 & 0.799  & 53.7  &  84.8 / 86.1  &  84.9 / 85.9  & 0.531  &  0.767 \\
    ALMT \cite{zhang2023learning} & 48.2 & 84.2 / 85.9  & 84.1 / 86.0 & 0.698 & 0.811 & 54.1 & 83.9 / 85.9 & 84.0 / 86.0  & 0.536 & 0.762 \\
    EMT$^\dag$ \cite{sun2024efficient} & 47.4 & 83.3 / 85.0 & 83.2 / 85.0 & 0.705 & 0.798 & 54.5 & 83.4 / 86.0 & 83.7 / 86.0 & 0.527 & 0.774 \\
    TMSON$^\dag$ \cite{xie2024trustworthy} & 47.4 & 85.4 / 87.2 & 85.4 / 87.2 & 0.687 & 0.809 & 55.6 & 85.2 / 86.4 & 85.3 / 86.2 & 0.526 & 0.766 \\
    EAU$^\dag$ \cite{guo2024embracing} & 48.8 & - / - & - / 86.2 & - & 0.809 & 54.8 & - / - & - / 86.9 & - & 0.816 \\
    KuDA$^\dag$ \cite{feng2024knowledge} & 47.1 & 84.4 / 86.4 & 84.5 / 86.5 & 0.705 & 0.795 & 52.9 & 83.3 / 86.5 & 83.0 / 86.6 & 0.529 & 0.776 \\
    MFON$^\dag$ \cite{zhang2025modal} & 44.9 & 84.8 / 86.9 & 84.8 / 86.9 & 0.725 & 0.797 & 53.7 & 82.7 / 86.3 & 83.1 / 86.3 & 0.528 & 0.780 \\
    DLF$^\dag$ \cite{wang2025dlf} & 47.1 & - / 85.1 & - / 85.0 & 0.731 & 0.781 & 53.9 & - / 85.4 & - / 85.3 & 0.536 & 0.764 \\
    MCL-MCF$^\dag$ \cite{fan2025multilevel} & - & 84.9 / 87.3 & 84.7 / 87.2 & 0.692 & 0.799 & - & 84.2 / 86.4 & 84.4 / 86.3 & 0.536 & 0.767 \\
    MLCL$^\dag$ \cite{zhuang2025mlcl} & 46.6 & 84.1 / 86.4 & 83.9 / 86.3 & 0.701 & 0.798 & 53.2 & 84.1 / 86.3 & 84.3 / 86.2 & 0.551 & 0.756 \\
    \midrule[0.5pt]
    MAG-XLNet\cite{} & 44.4 & - / 85.1 & - / 85.2 & 0.740 & 0.804 & 51.3 & - / 85.8 & - / 85.9 & 0.593 & 0.790\\
    MSG-MBA \cite{lin2023dynamically} & 46.1 & - / 87.0 & - / 87.0 & 0.696 & 0.816 & 52.6 & - / 86.3 & - / 86.3 & 0.581 & 0.795 \\
    UniMSE$^\dag$ \cite{hu2022unimse} & 48.7 & 85.9 / 86.9  & 85.8 / 86.4  & 0.691 & 0.809 & 54.4 & 85.9 / 87.5  & 85.8 / 87.5  & 0.523  & 0.773  \\
    MMML$^\dag$ \cite{wu2024mmml} & 48.3 & 85.9 / 88.2 & 85.9 / 88.2 & 0.643 & 0.838 & 55.0 & 86.3 / 86.7 & 86.2 / 86.5 & 0.517 & 0.791 \\
    ITHP$^\dag$ \cite{xiao2024neuroinspired} & - & - / 88.7  & - / 88.6  & 0.643 & 0.852 &  - & - / 87.3  & - / 87.4  & 0.564 & 0.813 \\
    \midrule[0.5pt]
    \rowcolor{gray!20} \textbf{MoB-BERT}  & 47.5 & 84.7 / 86.6 & 84.6 / 86.5  &  0.681 & 0.824  & 54.4 & 84.1 / 87.1  & 84.3 / 87.0  & 0.513 & 0.797 \\
    \rowcolor{gray!20} \textbf{MoB-RoBERTa}  & 49.9 & 86.6 / 88.9 & 86.5 / 88.8 & 0.612 & 0.854 &  \textbf{55.9} & 84.0 / 88.3 & 84.5 / 88.3 & 0.499 & 0.819 \\
    \rowcolor{gray!20} \textbf{MoB-DeBERTaV3}  & \textbf{51.1}  &  \textbf{87.4} / \textbf{89.5}  &  \textbf{87.3} / \textbf{89.5}  &  \textbf{0.598} & \textbf{0.873} & 55.1  &  \textbf{86.7} / \textbf{88.5}  &  \textbf{86.8} / \textbf{88.4}  &  \textbf{0.496} & \textbf{0.821}  \\
    \bottomrule[1.0pt]
\end{tabular}
}
\end{table*} 

\subsection{Implementation Details}
For language, on MOSI and MOSEI datasets, we employ three commonly-used language models trained on English scopes including BERT \cite{devlin2019bert}, RoBERTa \cite{liu2019roberta} and DeBERTaV3 \cite{he2021deberta,he2023debertav3}, while on SIMS and SIMSv2 datasets, we adopt three advanced language models tuned on Chinese scopes accordingly, including BERT-chinese \cite{devlin2019bert}, chinese-RoBERTa-wwm-ext \cite{cui2021wwm}, chinese-MacBERT \cite{cui2020macbert}. For acoustic and visual modalities, we adopt ImageBind \cite{girdhar2023imagebind} to extract the unimodal features for its excellent cross-modal alignment performance \cite{lin2024semanticmac}.


\subsection{Quantitative Results}
Comparing with current models, MoB achieves superior performance, outperforming previous methods on English (CMU-MOSI, CMU-MOSEI) and Chinese (CH-SIMS, CH-SIMSv2) benchmark datasets. Variants like MoB-DeBERTaV3 and MoB-RoBERTa achieve the best scores across a majority of key metrics. Besides, MoB consistently enhances the results of various pre-trained language models, revealing its excellent generalizability.

We attribute the effectiveness of MoB on diverse metrics to the decoupling of sentiment space with polarity (classification) and intensity (regression). This allows it to achieve better balance across two subtasks, simultaneously excelling at both where other models often compromise. The results also show a clear performance gain when using larger base models (DeBERTaV3) on English datasets, while on more challenging Chinese datasets, the improvements from model scaling (MacBERT) are more subtle, indicating the necessity of exploring cross-modal synergy.

\begin{table*}[htbp]
\caption{Performance comparison between the proposed MoB and baselines on CH-SIMS and CH-SIMSv2 datasets. The baseline models are reproduced except for $^\dag$ meaning that the corresponding results are copied from the original paper. }
\label{table_sims}
\centering
\scalebox{0.9}{
\begin{tabular}{c|cccccc|cccccc}
    \toprule[1.0pt]
    \multirow{2}{*}{Models} & \multicolumn{6}{c|}{CH-SIMS} & \multicolumn{6}{c}{CH-SIMS v2}\\
    & Acc5$\uparrow$  & Acc3$\uparrow$ & Acc2$\uparrow$ & F1$\uparrow$ & MAE$\downarrow$ & Corr$\uparrow$ & Acc5$\uparrow$  & Acc3$\uparrow$ & Acc2$\uparrow$ & F1$\uparrow$ & MAE$\downarrow$ & Corr$\uparrow$\\
    \midrule[1.0pt]
    Self-MM \cite{yu2021learning} & 43.8 & 66.1 & 79.3 & 79.4 & 0.416 & 0.600 & 53.5 & 72.7 & 78.7 & 78.6 & 0.315 & 0.691 \\
    MMIM \cite{han2021improving} & 43.3 & 66.8 & 78.4 & 78.1 & 0.431 & 0.587 & 50.5 & 70.4 & 77.8 & 77.8 & 0.339 & 0.641 \\
    AV-MC \cite{liu2022chsims2} & 45.5 & 68.5 & 79.7 & 80.2 & 0.372 & 0.685 & 52.1 & 73.2 & 80.6 & 80.7 & 0.301 & 0.721\\
    ALMT \cite{zhang2023learning} & 44.9 & 68.8 & 80.8 & 80.4 & 0.409 & 0.625 & 49.8 & 72.1 & 80.5 & 80.6 & 0.342 & 0.652 \\
    ConFEDE \cite{yang2023confede} & 42.5 & 69.4 & 81.6 & 81.4 & 0.391 & 0.669 & 50.2 & 71.6 & 80.4 & 80.2 & 0.355 & 0.702 \\
    EMT$^\dag$ \cite{sun2024efficient} & 43.5 & 67.4 & 80.1 & 80.1 & 0.396 & 0.623 & - & - & - & - & - & -\\
    TMSON$^\dag$ \cite{xie2024trustworthy} & 45.3 & 66.4 & 81.4 & 81.2 & 0.417 & 0.658 & - & - & - & - & - & -\\
    KuDA$^\dag$ \cite{feng2024knowledge} & 43.5 & 66.5 & 80.7 & 80.7 & 0.408 & 0.613 & 53.1 & \textbf{74.3} & 80.2 & 80.1 & \textbf{0.289} & 0.741 \\
    MFON$^\dag$ \cite{zhang2025modal} & - & - & 78.6 & 78.5 & 0.420 & 0.594 & - & - & - & - & - & -\\
    MCL-MCF$^\dag$ \cite{fan2025multilevel} & - & - & 81.8 & 81.8 & 0.410 & 0.588 & - & - & - & - & - & -\\
    MLCL$^\dag$ \cite{zhuang2025mlcl} & 45.5 & - & 81.4 & 81.3 & 0.412 & 0.596 & - & - & - & - & - & -\\
    \midrule[0.5pt] 
     \rowcolor{gray!20} \textbf{MoB-BERT}  & 45.1  & 69.2  &  81.2  &  81.2  &  0.381 & 0.696 &  51.2 &  72.7 & 81.3 & 81.2  & 0.336 & 0.727 \\
     \rowcolor{gray!20} \textbf{MoB-RoBERTa} & \textbf{45.7}  &  \textbf{70.9}  &  \textbf{81.9}  & \textbf{82.1}  & \textbf{0.371}  &  0.684  & 50.3  &  72.8  &  81.1  &  81.0  &  0.332  &  0.728  \\
    \rowcolor{gray!20} \textbf{MoB-MacBERT}  &  44.2   &  69.4 &  81.4  &  81.6 &  0.378  &  \textbf{0.697}  &  51.3  & 73.6  &  \textbf{82.1}  &  \textbf{82.1}  &  0.321  & \textbf{0.743}  \\
    \bottomrule[1.0pt]
\end{tabular}
}
\vspace{-0.2cm}
\end{table*} 

\begin{table}[htbp]
\centering
\setlength\tabcolsep{3.5pt}
\caption{Ablation study on the module of MoB on CMU-MOSI and CH-SIMS datasets. F1 is reported in positive/negative setting.}
\label{table_ablation}
\scalebox{0.9}{
    \begin{tabular}{ccccccccc}
    \toprule[1.0pt]
    \multirow{2}{*}{Module} & \multicolumn{4}{c}{CMU-MOSI} & \multicolumn{4}{c}{CH-SIMS} \\
    \cmidrule(r){2-5}\cmidrule(r){6-9}
    & Acc7$\uparrow$ & F1$\uparrow$ & MAE$\downarrow$ & Corr$\uparrow$ & Acc5$\uparrow$ & F1$\uparrow$ & MAE$\downarrow$ & Corr$\uparrow$ \\
    
    
    \midrule[1.0pt]
    \textbf{MoB}  & \textbf{51.1} & \textbf{89.5} &  \textbf{0.598} & \textbf{0.873} & \textbf{44.2} & \textbf{81.6} &  \textbf{0.378}  &  \textbf{0.697} \\
    \midrule[0.5pt]
    only IB &  47.1 & 87.1 & 0.652 & 0.849 & 40.6  &  77.2 & 0.436 & 0.594 \\
    only MoE &  46.5 & 86.5 & 0.671 & 0.835 & 38.7 & 76.8  & 0.425  &  0.586 \\
    \midrule[0.5pt]
    w/o PIBE & 49.4 & 87.7 &  0.641 & 0.859 & 41.2 & 79.0 & 0.418 & 0.631  \\
    w/o Polarity & 50.4 & 88.4 & 0.601 & 0.871 &  42.0 &  79.9 & 0.387  &  0.667 \\
    w/o Intensity & 47.4 &  89.1 & 0.631 & 0.863 & 42.4  & 80.6 & 0.389 &  0.669 \\
    w/o $\mathcal{L}_{ib}$ & 50.2 &  88.6 &  0.612 &  0.861 &  41.4 &  78.7 &  0.407 & 0.660 \\
    w/o $\mathcal{L}_{moe}$ &  49.2 & 89.4 & 0.602  & 0.871  & 43.9  &  81.4 & 0.379  & 0.683  \\
    w/o $\mathcal{L}_{btnk}$ & 47.7 & 88.7  & 0.623  & 0.858 &  43.0 &  79.4 &  0.398 & 0.659  \\
    \midrule[0.5pt]
    w/o MBRF & 48.3 & 87.4 & 0.626 & 0.862  &  41.3 & 78.5  & 0.406  &  0.645 \\
    w/o residual & 50.8 &  88.2 &  0.602 & 0.871  & 41.9  & 80.5  &  0.398 & 0.657 \\
    w/o $\mathcal{L}_{et}$ & 49.3 &  89.2 & 0.599 & 0.870  & 43.1 & 79.9 & 0.382 & 0.679 \\
    w/o $\mathcal{L}_{fusion}$ & 49.1 & 87.9  &  0.615 & 0.861  & 42.3 & 79.1 & 0.394 & 0.649 \\
    w/o $\mathcal{L}_{hard}$ & 48.6 & 88.9 &  0.609 &  0.854 &  40.9 & 79.0 &  0.412 &  0.638 \\
    \bottomrule[1.0pt]
\end{tabular}
}
\end{table}

\subsection{Ablation Study}

The ablation study demonstrates that MoB achieves its superior performance by effectively integrating three key innovations: the Information Bottleneck (IB) principle, Mixture-of-Experts (MoE) routing, and ordinal sentiment decoupling. Comparing the full MoB model against baselines using only IB or only MoE confirms that the synergistic combination of IB (for minimal yet sufficient information compression) and MoE (for dynamic specialization) is necessary for optimal results. Crucially, removing the specialized Polarity and Intensity Bottleneck Experts (PIBE) or Multimodal Bottleneck Routing Fusion (MBRF) results in the most significant performance degradations, further highlighting the vital role of both IB and MoE in learning compressed and task-relevant unimodal and multimodal features.

Additionally, the ordinal sentiment decoupling is proved highly effective, particularly the Intensity Experts component. Removing the intensity estimation module (w/o Intensity) cause substantial drops on most metrics, showing the importance in providing the fine-grained information needed for accurate sentiment prediction, especially for regression metrics like MAE and Corr. While the polarity recognition module focuses on the triple classes of sentiment, removing which (w/o Polarity) directly impacts the classification metrics including Acc7 and F1. The results on each module consistently validate that MoB is a tightly integrated system with PIBE and MBRF to capture intra- and inter-modal dynamics, governed by the bottleneck compressing, expert routing and hard mining losses.

\subsection{Flexible Modality input} 
As shown in Table \ref{table_modality}, the ablation study on diverse modality input demonstrate the MoB framework's exceptional flexibility and robustness in handling dynamic modal inputs. The model achieves optimal performance when all modalities ($\{l, a, v\}$) are available, effectively utilizing the dynamic bottleneck routing fusion to synthesize diverse cross-modal synergies. While when modalities are missing, MoB also exhibits strong adaptability. For instance, the performance drop is minimal when shifting from full-modal input to subsets containing language modality (e.g., $\{l, a\}$, $\{l, v\}$ or only $\{l\}$). This indicates that the routing mechanism can dynamically adjust its fusion strategy to effectively use available information, rather than relying on a static input structure. 

\begin{table}[htbp]
\centering
\setlength\tabcolsep{3.5pt}
\caption{Performance with flexible modality input for MoB on MOSI and SIMS datasets, where $\{l,a,v\}$ denote language, audio and vision modality, repectively. } 
\label{table_modality}
\scalebox{0.95}{
    \begin{tabular}{ccccccccc}
    \toprule[1.0pt]
    \multirow{2}{*}{Modality} & \multicolumn{4}{c}{CMU-MOSI} & \multicolumn{4}{c}{CH-SIMS} \\
    \cmidrule(r){2-5}\cmidrule(r){6-9}
    & Acc7$\uparrow$ & F1$\uparrow$ & MAE$\downarrow$ & Corr$\uparrow$ & Acc5$\uparrow$ & F1$\uparrow$ & MAE$\downarrow$ & Corr$\uparrow$ \\
    \midrule[1.0pt]
    $l,a,v$  & \textbf{51.1} & \textbf{89.5} &  \textbf{0.598} & \textbf{0.873} & 44.2 & 81.6 &  0.378  &  \textbf{0.697} \\
    $l,a$ & 51.0 &  89.3 & 0.599 &  0.870 & 43.7 & 80.1 &  0.401 & 0.643  \\
    $l,v$ & 49.9 & 89.4 & 0.602 &  0.867 &  43.2 & 80.5 & 0.412 &  0.639 \\
    $a,v$ & 20.6  & 55.7 & 1.446 & 0.207 & 36.5 & 70.1 & 0.475 & 0.516 \\
    only $l$ & 50.1 & 87.0 & 0.619 & 0.845 & 43.1  &  79.3 &  0.418 & 0.606  \\
    only $a$ & 19.7  & 54.9 & 1.516  &  0.113 &  26.4 & 63.1 & 0.539 & 0.358 \\
    only $v$ &  19.5 &  53.8 & 1.505 & 0.101 &  21.6 & 68.5 & 0.572 & 0.240 \\
    \bottomrule[1.0pt]
\end{tabular}
}
\end{table}

Furthermore, the result highlights the framework's capability to discern the contribution from each modality and dynamically assign appropriate bottleneck latents for sentiment prediction. The significant performance gap between "only $l$" and "only $a/v$" confirms that language modality plays the primary role for polarity recognition, while the inclusion of acoustic and visual modalities enhances the regression metric, suggesting their vital role in fine-grained intensity estimation. By applying the Information Bottleneck principle, MoB effectively filters out noise from inferior modalities (audio/vision) while extracting minimal yet sufficient latents to refine the final prediction.

\subsection{Variants of Router Settings} 
We evaluate the proposed two settings of bottleneck experts router in Table \ref{table_router}. Following \cite{han2024fusemoe}, we present three kinds of $Gate$ function (Softmax, Leplace, Gaussian) for the Multi-embed setting. Results demonstrate that on CMU-MOSI, the Multi-gate (MLP) setting achieves the highest classification metrics (Acc7: 51.1, F1: 89.5), while Multi-embed (Laplace) performs best in regression metrics (MAE: 0.581, Corr: 0.877). While on CH-SIMS, Multi-gate (MLP) excels on both classification and regression metrics (Acc5: 44.2, MAE: 0.378). Overall, Multi-gate (MLP) reaches its consistent excellence in the design of polarity and intensity router, which is attributed to its effectiveness in using learnable weights for dynamically assigning diverse latents based on the specific subtask. 

\begin{table}[htbp]
\centering
\setlength\tabcolsep{3.5pt}
\caption{Variants on router settings of MoB on CMU-MOSI and CH-SIMS datasets. F1 is reported in positive/negative setting.}
\label{table_router}
\scalebox{0.8}{
    \begin{tabular}{cccccccccc}
    \toprule[1.0pt]
    \multirow{2}{*}{Setting} & \multirow{2}{*}{Gate} & \multicolumn{4}{c}{CMU-MOSI} & \multicolumn{4}{c}{CH-SIMS} \\
    \cmidrule(r){3-6}\cmidrule(r){7-10}
    & & Acc7$\uparrow$ & F1$\uparrow$ & MAE$\downarrow$ & Corr$\uparrow$ & Acc5$\uparrow$ & F1$\uparrow$ & MAE$\downarrow$ & Corr$\uparrow$ \\
    
    \midrule[1.0pt]
    Multi-gate & $MLP$ & \textbf{51.1} & \textbf{89.5} &  0.598 & 0.873 & \textbf{44.2} & \textbf{81.6} &  \textbf{0.378}  &  \textbf{0.697} \\
    \midrule[0.5pt]
    \multirow{3}{*}{Multi-embed} 
    ~ & Softmax & 50.3 & 89.2 & 0.625 & 0.853  & 42.6  & 80.1  & 0.411 & 0.663  \\
    ~ & Laplace & 51.0 & 89.4 & \textbf{0.581} & \textbf{0.877} & 43.5 & 80.9 &  0.412 & 0.669 \\
    ~ & Gaussian & 49.9  &  88.9 &  0.597 &  0.871 & 43.9  & 81.2 & 0.382 &  0.683 \\
    \bottomrule[1.0pt]
    
\end{tabular}
}
\end{table}

\subsection{Polarity and Intensity Routing During Training}

\begin{figure*}[htbp]
    \centering
    \subfigure[Language Modality] {
     \label{vis_lang_routing}
     \begin{minipage}[b]{0.31\linewidth}
        \rotatebox{90}{\scriptsize{~~~~~~~~~~~~~~~~~Polarity}} 
        \centering
        \includegraphics[width=0.91\linewidth]{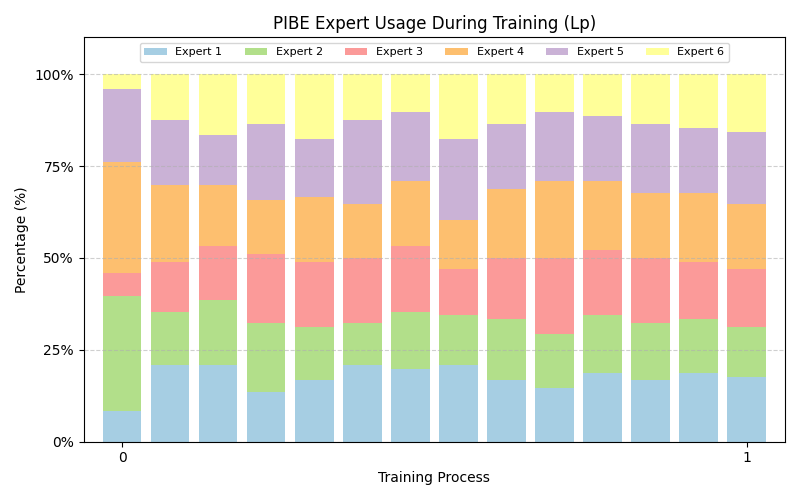}\vspace{2pt}
        \rotatebox{90}{\scriptsize{~~~~~~~~~~~~~~~~~Intensity}} 
        \includegraphics[width=0.91\linewidth]{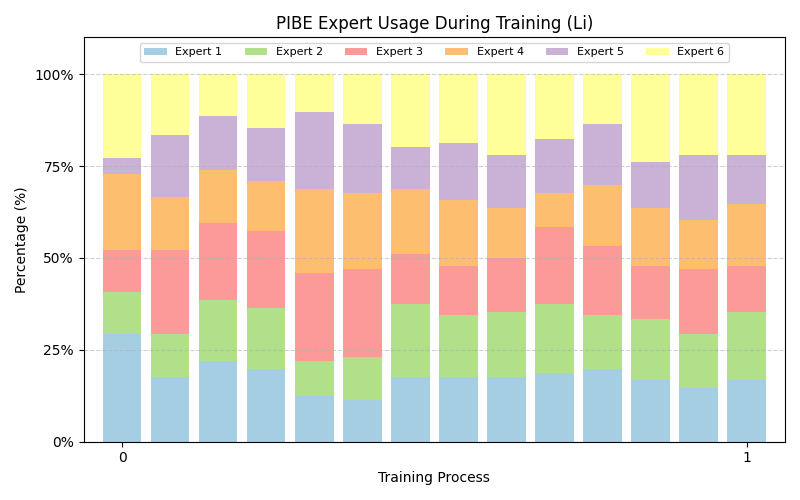}
    \end{minipage}}
    \subfigure[Audio Modality] { 
    \label{vis_audio_routing}
     \begin{minipage}[b]{0.31\linewidth}
        \centering
        \includegraphics[width=0.91\columnwidth]{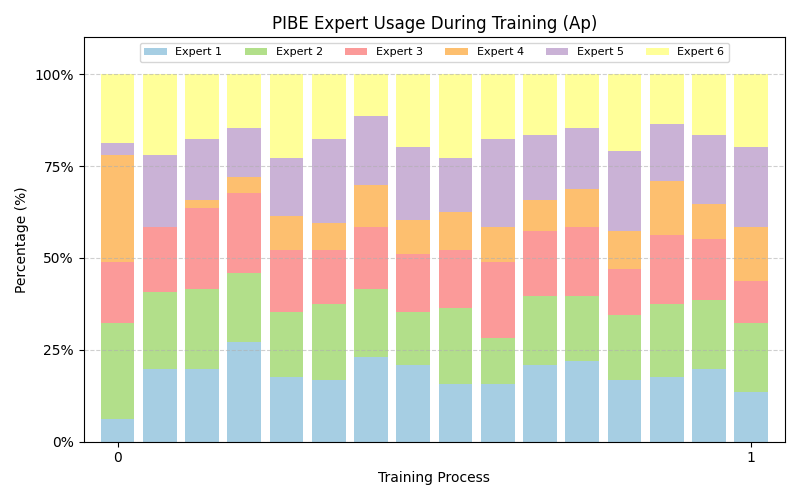}\vspace{2pt}
        \includegraphics[width=0.91\columnwidth]{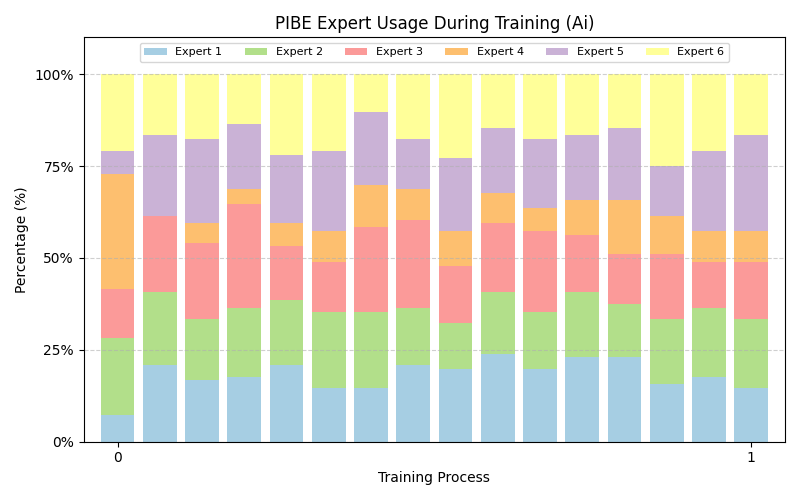}     
    \end{minipage}}
    \subfigure[Vision Modality] {
     \label{vis_vision_routing}
     \begin{minipage}[b]{0.31\linewidth}
        \centering
        \includegraphics[width=0.91\columnwidth]{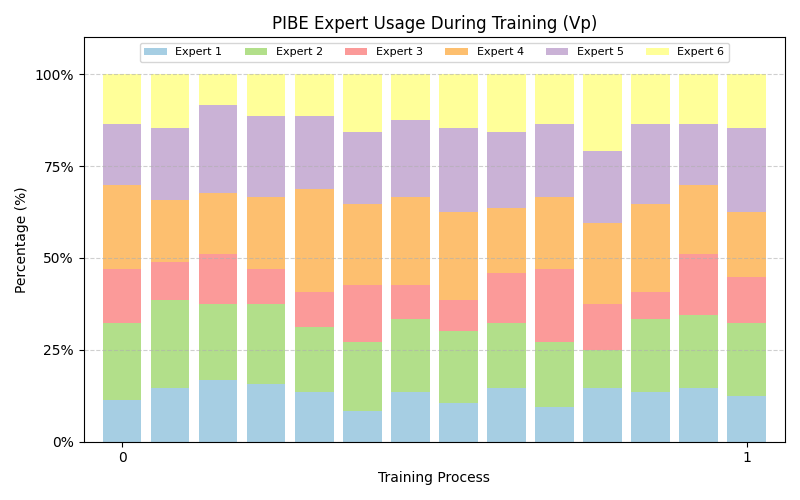}\vspace{2pt}
        \includegraphics[width=0.91\columnwidth]{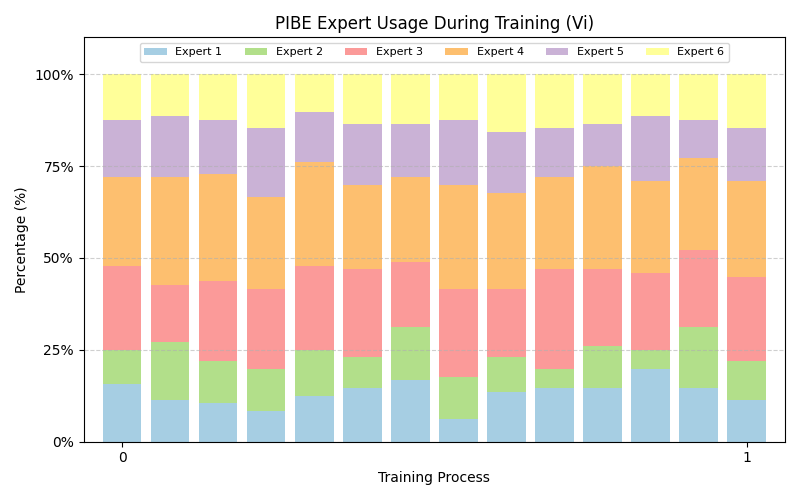}  
    \end{minipage}}
    \caption{Tracing of the polarity and intensity routing during training for (a) language, (b) audio and (c) vision modality. }
    \label{vis_pibe_routing}
\end{figure*}

As shown in Figure \ref{vis_pibe_routing}, the visualization of expert usage during training confirms the highly effective and dynamic routing capabilities of the MoB framework across language, audio, and vision modalities for either Polarity or Intensity tasks. A key observation is the consistent and near-uniform utilization of all six bottleneck experts on both subtask for each modality after efficient training. The contribution of experts shift from imbalance to balance during training, which demonstrates the effectiveness of MoB router and ensures every expert to actively involve in capturing the necessary fine-grained latents required for accurate prediction.

Moreover, the consistent and balanced expert utilization across all three modalities strongly suggests MoB possesses excellent generalization ability in dynamically handling diverse input data. The framework doesn't rely on one modality for each subtask. Such uniform distribution indicates that the router can robustly identify and route task-relevant information, enabling the model to learn better modality-specific representations. The balanced engagement across the multimodal input space is a critical factor in preventing performance collapse when one modality is noisy or missing, thereby promoting overall model robustness and generalization.

\begin{figure}[htbp]
\centering 
\includegraphics[scale=0.4]{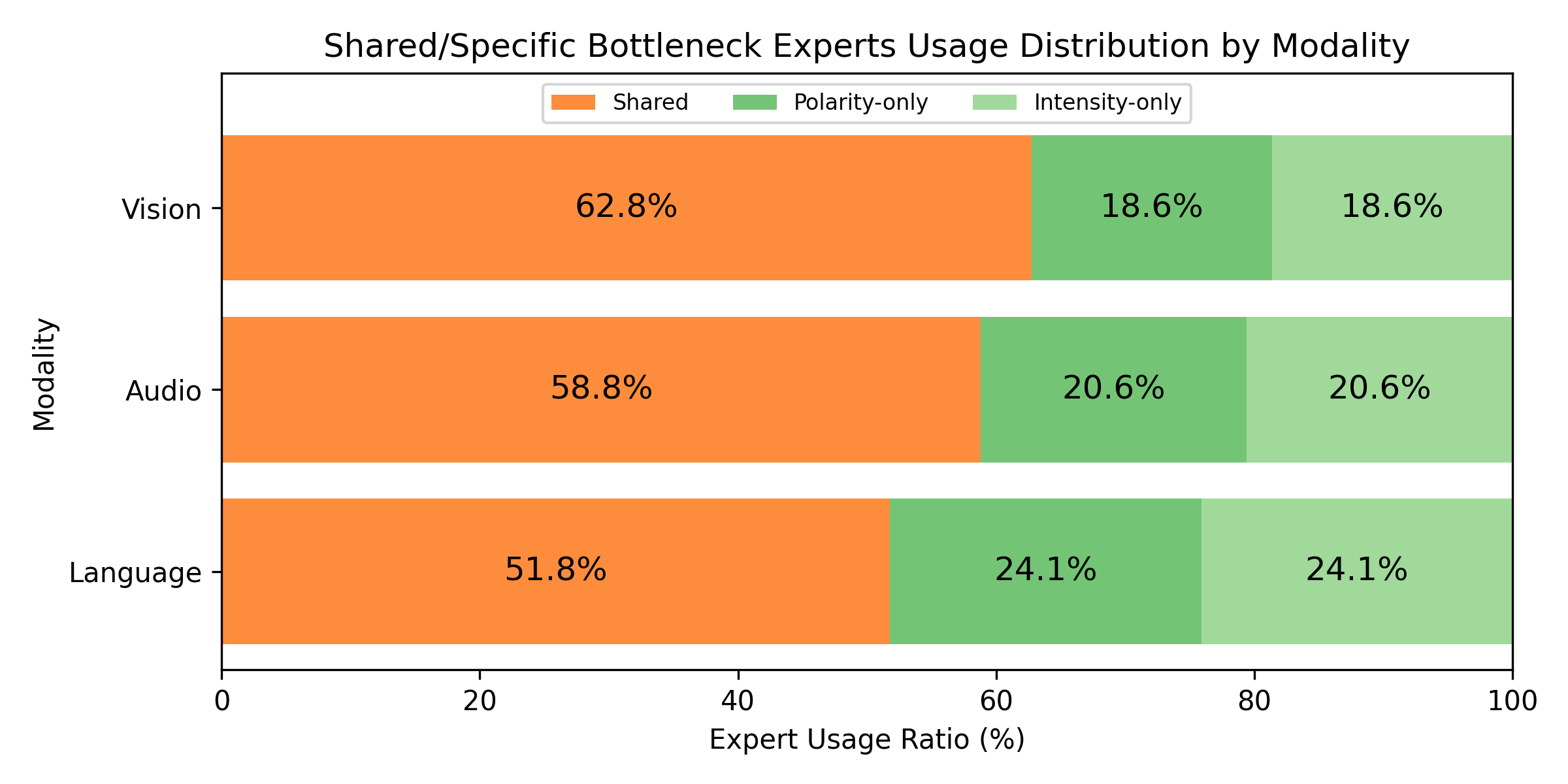}
\caption{Illustration of the shared and specific expert usage ratio between the polarity and intensity router on MOSI. }
\label{vis_share_specific}
\end{figure}

\subsection{Task-aware Routing During Inference.} 
The visualization in Table \ref{vis_share_specific} provides the shared and specific expert usage ratio during inference, explicitly demonstrating the effectiveness of PIBE routers in decoupling and assigning task-aware latents. By dynamically categorizing the bottleneck experts into "Shared", "Polarity-only" and "Intensity-only", the result proves MoB does not blindly extract information but actively discerns each latent according to specific subtask, revealing trustworthy unimodal feature extraction.

We observe that the dominance of the "Shared" category across all three modalities (ranging from roughly 51.8$\%$ to 62.8$\%$) illustrates that the intrinsic correlation between polarity recognition and intensity regression subtasks  highly rely on shared information. This aligns with the natural structure of sentiment, where the emotional sign (polarity) and its magnitude (intensity) often stem from the same core expressions, such as a smile or a distinct tonal shift. Besides, the existence of distinct "Polarity-only" and "Intensity-only" parts confirms that MoB effectively avoids feature redundancy in each subtask, isolating unique cues that would be otherwise confused in a purely joint latent space.

Compared to vision (62.8$\%$ shared) and audio (58.8$\%$ shared), language has the lowest shared ratio (51.8$\%$) and the highest proportion of task-specific experts (24.1$\%$ for each). This indicates that the router assigns more fine-grained and distinctive features for language modality, leveraging its rich semantic capacity in explicitly decouple sentiment polarity from strength. In contrast, audio and vision modalities rely more on holistic and shared representations, suggesting that non-verbal cues tend to contribute to polarity and intensity simultaneously rather than independently.

\subsection{Dynamics in Bottleneck Routing Fusion} 
\begin{figure}[htbp]
\centering 
\includegraphics[scale=0.45]{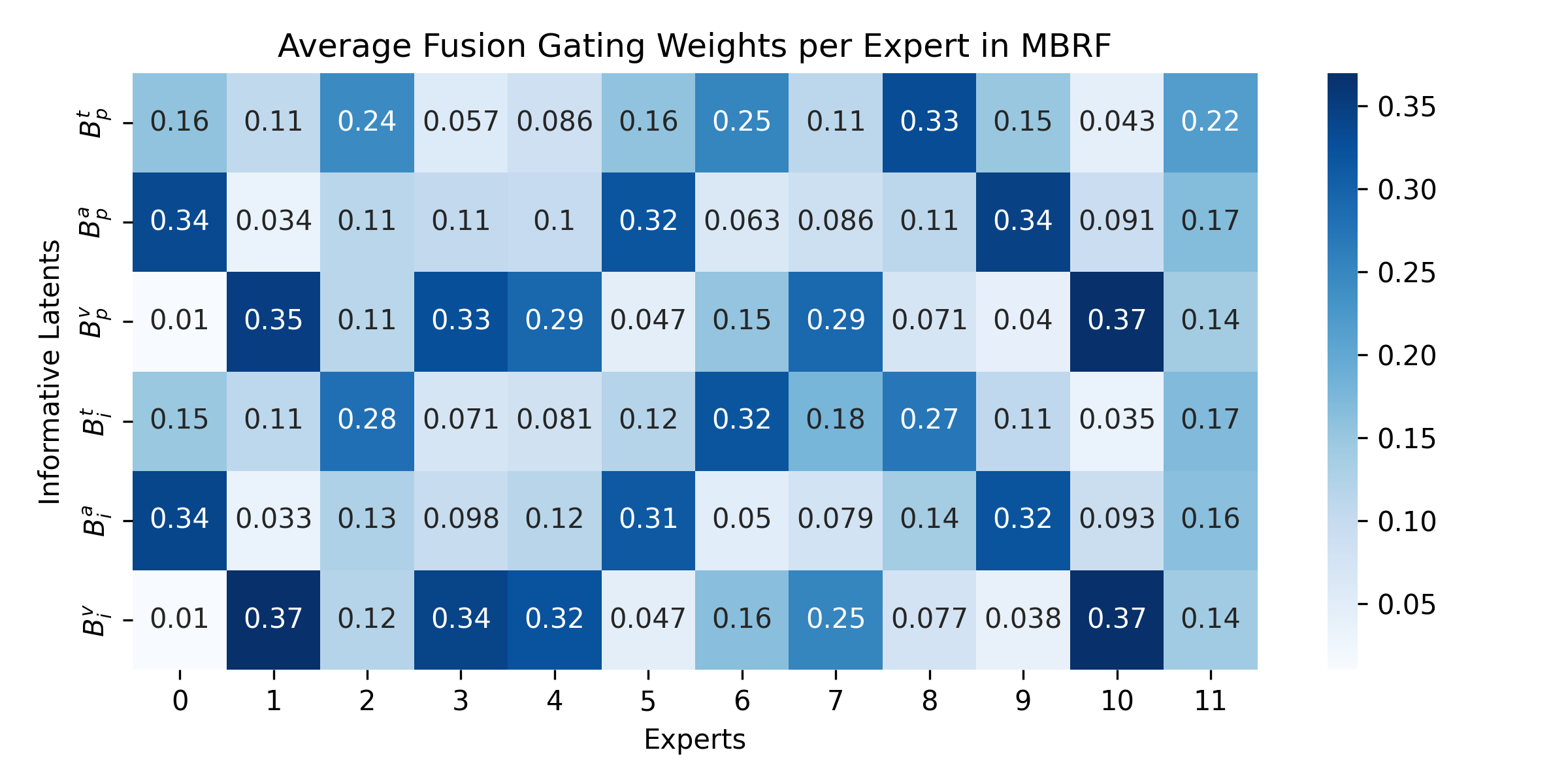}
\caption{Visualization on the gating weight of informative unimodal latents routing during training for multimodal bottleneck routing fusion module.. }
\label{vis_mbrf_routing}
\end{figure}

We visualize the gating weights of the experts in MBRF in Figure \ref{vis_mbrf_routing}, indicating MoB's  sophisticated capacity in managing cross-modal interactions by dynamically assigning diverse experts to model different types of information flows, ranging from self-modal preservation to complex multimodal fusion. Instead of integrating features from different modalities jointly, which often leads to noise and interference due to modality gap, the router specializes experts to handle specific interaction pathways. This decoupling pattern confirms that the framework effectively minimizes mutual interference among diverse cross-modal synergy, which ensures that the unique characteristics of each modality are preserved while simultaneously enabling effective fusion process.

From Figure \ref{vis_mbrf_routing}, we can observe that some experts are dedicated to refining unimodal signals, while others are responsible for capturing the synergy between modalities, illustrating the fine-grained specialization in distinguishing unimodal and cross-modal dynamics. For instance, Expert 8 acts as a clear language specialist, assigning dominant weights to language modality ($B_p^l$ 0.33, $B_i^l$ 0.27) while suppressing audio and visual inputs, thereby safeguarding the semantic core for textual information. Similarly, Expert 9 focuses heavily on acoustic cues ($B_p^a$ 0.34, $B_i^a$ 0.32), isolating the acoustic information. In contrast, the router assigns Expert 11 to handle complex relationship between facial expressions and voice tone, as indicated by its simultaneous high weights on both visual ($B_p^v$ 0.14, $B_i^v$ 0.14) and audio ($B_p^a$ 0.17, $B_i^a$ 0.16) latents. This proves that the MBRF automatically learns to allocate specific experts to capture the interaction across verbal and non-verbal signals, distinguishing cross-modal synergy from independent unimodal features.

Ultimately, the observed patterns of highly specialized yet balanced expert engagement across modalities confirm that the Information Bottleneck principle, when integrated with Mixture-of-Experts routing, successfully achieving the desired dynamic multimodal learning scheme for interpretability and adaptability of video-based MSA.

\section{Conclusion}
In this paper, we present Mixture-of-Bottleneck (MoB) framework, a novel approach for dynamic and comprehensive video-based multimodal sentiment analysis. By reformulating sentiment into an ordinal space, MoB explicitly decouples polarity recognition and intensity estimation subtasks. With integration of the IB and MoE routing, MoB enables each expert to specialize in extracting compact and task-relevant information from different modalities. The designed multimodal bottleneck routing fusion further captures fine-grained cross-modal synergy adaptively. Extensive experiments validate its superior performance and better interpretability.


\bibliographystyle{IEEEtran}
\bibliography{MSA_ref}




\end{document}